\documentclass[conference]{IEEEtran}
\IEEEoverridecommandlockouts

\usepackage{cite}
\usepackage{amsmath,amssymb,amsfonts}
\usepackage{graphicx}
\usepackage{textcomp}
\usepackage{xcolor}
\def\BibTeX{{\rm B\kern-.05em{\sc i\kern-.025em b}\kern-.08em
    T\kern-.1667em\lower.7ex\hbox{E}\kern-.125emX}}

\usepackage{algorithm}
\usepackage{algorithmic}
\usepackage{array}     
\usepackage{tabularx}  
\usepackage{hyperref}
\usepackage{subcaption}
\usepackage{multirow}
\usepackage{soul}
\usepackage{orcidlink}

\begin{document}

\title{An Efficient Privacy-preserving Intrusion Detection Scheme for UAV Swarm Networks 
\thanks{
This paper has been accepted for publication in the Proceedings of the 44th AIAA/IEEE Digital Avionics Systems Conference (DASC) 2025, where it received the Best Paper of Session Award. \\
The source code for this work is available at: \url{https://github.com/SPIRE-Lab-2025/UAV-IDS-FL}}
}

\author{\IEEEauthorblockN{Kanchon Gharami \orcidlink{0000-0003-0032-8201}}
\IEEEauthorblockA{\textit{Dept. of Electrical Engineering and Computer Science} \\
\textit{Embry-Riddle Aeronautical University, Florida, USA}\\
gharamik@my.erau.edu, kanchon2199@gmail.com}
\and
\IEEEauthorblockN{Shafika Showkat Moni \orcidlink{0000-0002-7710-4217}}
\IEEEauthorblockA{\textit{Dept. of Electrical Engineering and Computer Science} \\
\textit{Embry-Riddle Aeronautical University, Florida, USA}\\
monis@erau.edu, shafika1403@gmail.com}
}

\maketitle
\begin{abstract}
The rapid proliferation of unmanned aerial vehicles (UAVs) and their applications in diverse domains, such as surveillance, disaster management, agriculture, and defense, have revolutionized modern technology. While the potential benefits of swarm-based UAV networks are growing significantly, they are vulnerable to various security attacks that can jeopardize the overall mission success by degrading their performance, disrupting decision-making, and compromising the trajectory planning process. The Intrusion Detection System (IDS) plays a vital role in identifying potential security attacks to ensure the secure operation of UAV swarm networks. However, conventional IDS primarily focuses on binary classification with resource-intensive neural networks and faces challenges, including latency, privacy breaches, increased performance overhead, and model drift. This research aims to address these challenges by developing a novel lightweight and federated continuous learning-based IDS scheme. Our proposed model facilitates decentralized training across diverse UAV swarms to ensure data heterogeneity and privacy. The performance evaluation of our model demonstrates significant improvements, with classification accuracies of 99.45\% on UKM-IDS, 99.99\% on UAV-IDS, 96.85\% on TLM-UAV dataset, and 98.05\% on Cyber-Physical datasets.

\end{abstract}

\begin{IEEEkeywords}
UAV, Intrusion detection, UAV swarm, Federated learning, Cybersecurity, Anomaly detection, Heterogeneous learning, Privacy-preserving
\end{IEEEkeywords}

\section{Introduction}
\IEEEPARstart{U}{nmanned} Aerial Vehicles (UAVs) have emerged as transformative tools across a wide range of fields, including disaster response, logistics, precision agriculture, environmental monitoring, and military operations~\cite{mohammed2023comprehensive, al2024machine}. Their ability to perform tasks autonomously in inaccessible or hazardous areas, combined with real-time adaptability and scalability, makes them indispensable for critical applications. However, the growing reliance on UAVs introduces significant security challenges, especially in swarm networks where multiple UAVs collaborate to achieve shared objectives. In swarm-based systems, where decisions and actions are interdependent, a single compromised UAV can cascade and compromise the entire network. Intrusion Detection Systems (IDSs) play a vital role in safeguarding UAV swarm networks by identifying and mitigating threats in real-time. While traditional IDSs have demonstrated effectiveness in static or terrestrial networks, UAV swarm environments introduce unique challenges, such as limited computational resources, high mobility, and the need for decentralized decision-making. To address these challenges, researchers are focusing on developing advanced IDSs tailored to the dynamic and resource-constrained nature of UAV swarm networks. Intrusion detection systems (IDSs) can mitigate the potential risk of these security vulnerabilities.

Despite notable progress in swarm-based IDS for UAVs, unexplored avenues remain. Firstly, they lack support for handling heterogeneous data and require manual tuning to adapt to new datasets, at least to match the dataset dimension. Developing a unified framework that can seamlessly adapt to diverse datasets without changes to model architecture or hyperparameters while maintaining high accuracy is crucial for UAV swarm networks. Secondly, studies utilizing resource-intensive deep learning models make it difficult for deployment in resource-constraint UAV swarm networks. Thirdly, conventional IDS encounters difficulties in enabling the model to retain knowledge from earlier datasets while seamlessly learning from new data or swarm groups. Thus, this results in a decline in the accuracy of identifying intrusion and a potential increase in false positives or false negatives. Finally, the existing model designed under controlled laboratory conditions trained on any particular dataset does not reflect all the vast complexities of real-world scenarios.

We aim to overcome the above challenges that conventional intrusion detection models face including latency, privacy breaches, increased performance overhead, and model drift. We propose a lightweight and efficient multiclass intrusion detection framework for UAV swarm networks, leveraging federated continuous learning~\cite{li2020review, yang2024federated}. Unlike existing research that focuses on one specific dataset, we introduce a unified three-component architecture to support data heterogeneity in our IDS model. Firstly, the swarm-specific input layer is designed to accommodate the heterogeneity of UAV swarm datasets, where the number of input features may differ across swarms due to varying sensor configurations or attack types. This layer transforms datasets with varying feature dimensions into a fixed-dimensional representation, ensuring compatibility with subsequent components and federated learning (FL)~\cite{wen2023survey}. It consists of a dense layer followed by a custom activation function, ReTeLU, which combines the noise-filtering properties of ReLU with the selective thresholding of TeLU. This hybrid activation effectively eliminates irrelevant noise while retaining significant features, reducing the impact of minor sensor fluctuations common in UAV data. As a result, the model can focus on meaningful and stable patterns, enhancing its robustness in dynamic environments.

Secondly, the shared CNN-LSTM encoder serves as the backbone of the architecture, featuring a parallel setup where the same input data is processed simultaneously by Convolutional Neural Network (CNN)~\cite{priyadarsini2025cnn} and Long Short-Term Memory (LSTM)~\cite{yu2019review}. This design allows the model to capture spatial and temporal patterns concurrently, making it highly efficient for intrusion detection in resource-constrained UAV environments. Additionally, this encoder also participates in federated learning, enabling collaborative knowledge sharing across heterogeneous swarms. Then, by integrating Elastic Weight Consolidation (EWC) into the FL setup, we transform it into a Federated Continuous Learning system to enable the model to retain knowledge from earlier datasets while seamlessly learning from new data or swarm groups. Moreover, adapting FL, where data remain securely stored on local devices and only aggregated model updates are communicated, facilitates the development of a privacy-preserving IDS model.

Finally, the swarm-specific classifier layer acts as a multi-head classifier, tailored to the unique classification demands of each swarm group. This layer consists of a lightweight two-layer neural network and employs our custom ReTeLU activation function, ensuring efficient and swarm-specific anomaly detection.

\subsection{Related Works}
In recent years, a wide range of methods have been developed to tackle the challenge of intrusion detection in UAV networks. This section provides an overview of recent approaches focusing on intrusion detection, analyzing their strengths, and identifying their limitations, which serve as a foundation for developing our proposed framework.

In 2019, Bithas et al.~\cite{bithas2019survey} conducted an in-depth assessment of UAV anomaly detection using deep learning techniques. They evaluated four deep learning architectures: autoencoders, Deep Belief Networks (DBN), and Long Short-Term Memory (LSTM) networks. Their study used several datasets, including KDD Cup 1999~\cite{tavallaee2009detailed}, NSL-KDD~\cite{bala2019review}, CICIDS2017~\cite{panigrahi2018detailed}, and CICIDS2018~\cite{cantonegeneralization}. A notable strength of this research is its approach to addressing challenges such as channel modeling and resource management. However, a significant limitation is the reliance on outdated datasets like KDD Cup 1999 and NSL-KDD, which may not reflect the complexities of modern UAV systems.

Raju et al.~\cite{dhakal2023uav}, in 2023, conducted a UAV anomaly detection study using a Variational Autoencoder (VAE) on the ALFA dataset~\cite{keipour2021alfa}. Their work employs unsupervised learning, which enhances the system’s ability to detect unseen attacks within a lightweight model and to cluster various attack types into groups. However, this study has limitations, including an inability to explicitly identify the underlying causes of detected anomalies and a potential data bias due to over-reliance on a single dataset.


In 2024, Hadi et al.~\cite{hadi2024real} introduced a collaborative intrusion detection framework using a feedforward network. Their approach emphasizes collaborative detection through a trust score mechanism rather than relying solely on deep learning models, offering a new perspective on UAV intrusion detection. Additionally, they combine ReLU and TeLU activation functions to handle noisy sensor data effectively. The framework achieves high detection accuracy on the UAV\_IDS\cite{UAVIDS2020_Dataset} dataset, even in real-world deployments. However, their work lacks support for handling heterogeneous data and swarm networks, privacy preservation, and multiclass classification. Furthermore, while their system has been deployed in real-world UAV test runs, the experiments were still conducted within a limited testbed, which may not fully capture the complexity of real-life attack patterns.

Zhao et al.~\cite{zhao2024security} propose a security situation assessment approach for UAV swarm networks using a Transformer-ResNeXt-SE-based model, one of the most advanced neural network architectures among all the anomaly detection literature. Their focus is on providing proactive network protection rather than detecting anomalies post-attack. The model is exceptionally well-designed, and the collaborative detection strategy is logically sound, yielding strong evaluation results on four test datasets. However, the main limitation of this system is its heavy computational requirements, making it unsuitable for resource-constrained UAVs. In addition to that, the lack of support for handling heterogeneous data could introduce system performance biases.

Most recently, Lu et al.~\cite{lu2024swarm} proposed a swarm anomaly detection model for IoT UAVs using a multimodal denoising autoencoder. They successfully implemented their model in a simulated federated learning setup, ensuring privacy preservation. The denoising autoencoder effectively detects multiclass attacks, even in the presence of noisy sensor data. Their experiments demonstrate outstanding performance across four advanced intrusion detection datasets. However, the model has some limitations: it is designed for homogeneous swarm groups (with homogeneous datasets), and by adding Gaussian noise to the sensor data~\cite{guo2011estimation}, there is a noticeable drop in detection accuracy on the Cyber\_Physical dataset, which contains both sequential cyber data and non-linear sensor data.

All of the above research has significantly contributed to the UAV intrusion detection field. Through a review and analysis of the literature, it is clear that while traditional UAV intrusion detection models have unique strengths, none serve as a unified IDS framework. To the best of our knowledge, none of the existing methodologies addresses data heterogeneity, which is crucial for developing generalized decentralized ad hoc UAV networks. 

\subsection{Contributions}
This paper aim to address the above research gaps in existing IDS for UAV swam networks. We propose a unified multiclass intrusion detection framework leveraging federated continuous learning. Our system's core is an encoder-classifier network built on a CNN-LSTM multimodal architecture~\cite{praharaj2023hierarchical}. By combining the ability of the CNN to handle non-linear data with the strength of LSTM to process time series information, our model efficiently handles intrusion detection data sets that include both cyber attributes (linear data types) and physical attributes (non-linear data types). Our framework adopts a federated continuous learning approach~\cite{li2020review, yang2024federated}, enabling the system to train simultaneously on heterogeneous datasets while preserving data privacy~\cite{ilhan2023scalefl}. We test and train the model across the four distinct swarms for our four heterogeneous datasets: UAV-IDS, UKM-IDS, TLM-IDS, and Cyber-Physical. We summarize the key contributions of this paper as follows:

\begin{itemize}
 \item The proposed framework integrates federated learning to collaboratively train on four heterogeneous datasets, effectively reducing bias toward any specific dataset or environment, and improving generalization. The modular and scalable design supports the seamless addition of new swarm groups with new datasets and attributes, making the learning process more generalized and adaptive.
  
  \item Training on heterogeneous datasets strengthens the model's ability to recognize anomaly patterns, which we leverage to detect attacks early and effectively through binary classification in our supervised model.
  
  \item To address the challenge of limited perspective, we integrate the Elastic Weight Consolidation (EWC)~\cite{thorne2020elastic} regularization technique into our FL setup, transforming it into a Federated Continuous Learning system. This integration prevents catastrophic forgetting~\cite{kirkpatrick2017overcoming}, enabling the model to retain knowledge from earlier datasets while seamlessly learning from new data or swarm groups.

  \item Our lightweight Encoder-Classifier network incorporates a shared CNN-LSTM encoder to efficiently process heterogeneous datasets, while the multilayer perceptron (MLP) classifier~\cite{gardner1998artificial} dimensions vary across swarms to enable multiclass classification. This architecture minimizes hardware requirements while supporting diverse swarm groups with different class label sets.
  
  \item FL's decentralized approach ensures that each swarm group retains its data locally and trains through its edge server, eliminating the need to transmit data to the Global Control Server (GCS). This approach guarantees robust data privacy for every swarm group.
  \end{itemize}
The rest of the paper is organized as follows: Section II outlines the proposed Encoder-Classifier model. We discuss the implementation details, including datasets, experimental setups, and federated learning integration in Section III. Section IV presents the results and analysis, including performance metrics and comparative benchmarks. Finally, Section V concludes with a summary of findings.

\section{Proposed Encoder-Classifier Model}
\begin{figure*}[ht]
    \centering
    \includegraphics[width=0.8\textwidth]{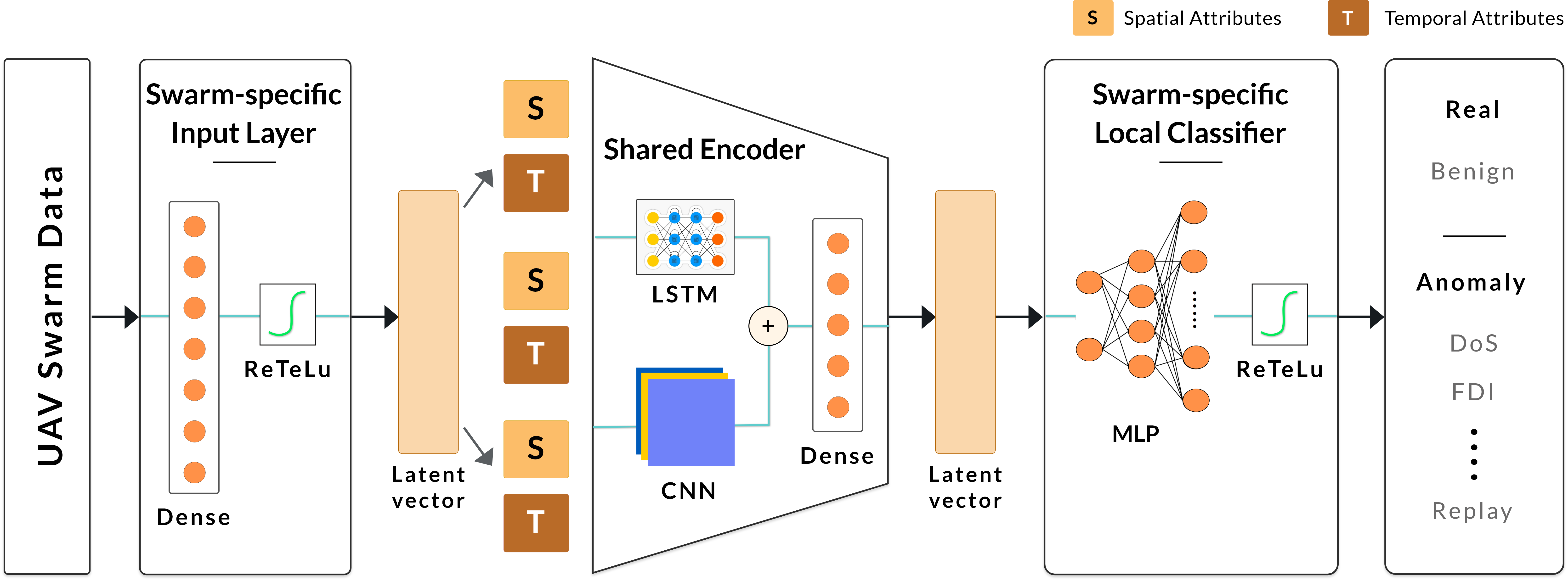}
    \caption{The model diagram of the proposed Encoder-Classifier framework.}
\label{fig:model_diagram}.
\end{figure*}

The proposed Encoder-Classifier model combines CNN for spatial feature extraction with LSTM networks to capture temporal dependencies, enabling efficient multiclass classification. It is designed to handle the heterogeneity of UAV swarm datasets, which can vary in both input features and attack class labels. To address this, the model includes a swarm-specific input layer that adapts to the varying input dimensions of each dataset across different swarm, along with a local classifier that adjusts its output based on the number of attack classes for each swarm. As shown in Fig.~\ref{fig:model_diagram}, the CNN-LSTM encoder extracts a unified latent representation from the input, while the MLP classifier tailors the final classification output to match the specific needs of each swarm. This structure ensures that the model can perform multiclass intrusion detection across different UAV networks, maintaining flexibility and accuracy in diverse environments.

\subsection{Swarm-Specific Input Layer}

The swarm-specific input layer is designed to accommodate the heterogeneity of UAV swarm datasets, where the number of input features may differ across swarms due to varying sensor configurations or attack types. This layer adapts each swarm's dataset to a fixed-dimensional representation, ensuring consistency and compatibility for federated learning with the rest of the model, regardless of differences in input size.

Given an input dataset \( \mathbf{X}_i \in \mathbb{R}^{B \times F_i} \), where \(B\) is the batch size and \(F_i\) is the number of features for the \(i\)-th swarm, we project the data to a fixed hidden dimension \( \mathbf{h}_i \in \mathbb{R}^{B \times D} \) using a fully connected layer:
\begin{equation}
\mathbf{h}_i = \text{ReTeLU}\left( \sum_{j=1}^{F_i} \mathbf{W}_{ij} \mathbf{X}_{ij} + \mathbf{b}_i \right)
\end{equation}

where\( \mathbf{W}_{ij} \) and \( \mathbf{b}_i \) are the weight matrix and bias, respectively, and ReTeLU is a customized activation function. The ReTeLU function combines ReLU and TeLU (Thresholded ReLU) to enhance non-linearity and improve the model’s robustness to noisy data. Specifically, ReLU sets negative values to zero, eliminating irrelevant noise \cite{hadi2024real}, while TeLU introduces a threshold \( \theta \), ensuring only the most significant positive features are retained, like following: 
\begin{equation}
\text{ReTeLU}(x) = \begin{cases}
0, & \text{if } x \leq 0 \\
\theta, & \text{if } 0 < x \leq \theta \\
x, & \text{if } x > \theta
\end{cases}
\end{equation}
This hybrid activation reduces the impact of minor sensor fluctuations, which are common in UAV data, allowing the model to focus on more meaningful and stable patterns.

The Client-Specific Input Layer operates independently within each swarm, meaning its weights are updated locally and do not participate in federated learning aggregation. This localized adaptation ensures that each swarm can process its unique dataset effectively without requiring synchronization with other swarms. The fixed output dimension of this layer, regardless of the input variability, maintains consistency across swarms, ensuring that the model’s latent space is uniform. This enables the model to integrate data from heterogeneous swarms and perform multiclass classification effectively, leveraging federated learning while preserving each swarm's specific characteristics.

\subsection{Shared CNN-LSTM Encoder}
The shared CNN-LSTM encoder combines CNN for spatial feature extraction and LSTM networks for capturing temporal dependencies~\cite{kuettel2010s}, making it ideal for intrusion detection with resource-limited UAV hardware. Given the input \( \mathbf{h} \in \mathbb{R}^{B \times D} \), where \(B\) is the batch size and \(D\) is the hidden dimension, the encoder first processes the data through two parallel layers: CNN and LSTM.

The CNN path extracts spatial features from the input data by applying a convolution operation, where \( \mathbf{W}_{\text{cnn}} \in \mathbb{R}^{D \times C_{\text{cnn}}} \) is the weight matrix and \( \mathbf{b}_{\text{cnn}} \in \mathbb{R}^{C_{\text{cnn}}} \) is the bias term. The output of the convolution is then passed through the ReTeLU activation function (for each filter k) and followed by a max-pooling layer to reduce the sequence length and retain the most important features. This process is represented by:
\begin{align}
\mathbf{x}_{\text{cnn}, k} &= \text{ReTeLU} \left( \sum_{i=1}^{D} \mathbf{W}_{\text{cnn}, k, i} \mathbf{h}_i + \mathbf{b}_{\text{cnn}, k} \right) \\
\mathbf{x}_{\text{cnn}}' &= \text{MaxPool1D} \left( \mathbf{x}_{\text{cnn}} \right)
\end{align}
Meanwhile, the LSTM path processes the data sequentially to capture temporal dependencies. The LSTM operations are given by:
\begin{align}
\tilde{\mathbf{f}}_t &= \sigma\left( \mathbf{W}_f \cdot \mathbf{z}_t + \mathbf{b}_f \right) \\
\tilde{\mathbf{i}}_t &= \sigma\left( \mathbf{W}_i \cdot \mathbf{z}_t + \mathbf{b}_i \right) \\
\tilde{\mathbf{o}}_t &= \sigma\left( \mathbf{W}_o \cdot \mathbf{z}_t + \mathbf{b}_o \right) \\
\tilde{\mathbf{c}}_t &= \tilde{\mathbf{f}}_t \circ \tilde{\mathbf{c}}_{t-1} + \tilde{\mathbf{i}}_t \circ \tanh\left( \mathbf{W}_c \cdot \mathbf{z}_t + \mathbf{b}_c \right) \\
\tilde{\mathbf{h}}_t &= \tilde{\mathbf{o}}_t \circ \tanh(\tilde{\mathbf{c}}_t)
\end{align}


Where, $\mathbf{z}_t = [\tilde{\mathbf{h}}_{t-1}; \mathbf{x}_t]$. The final output from the LSTM is the last hidden state, \( \mathbf{h}_{\text{lstm}} = \mathbf{h}_t \). After both the CNN and LSTM paths process the data, their outputs are concatenated into a single feature vector, \( \mathbf{z} \in \mathbb{R}^{B \times (C_{\text{cnn}} + L_{\text{lstm}})} \), where \( C_{\text{cnn}} \) is the number of channels after the CNN layer and \( L_{\text{lstm}} \) is the number of LSTM features. This concatenated vector is then passed through a fully connected layer to project the combined features into the latent space:
\begin{align}
\mathbf{z} &= \text{concat} \left( \mathbf{x}_{\text{cnn}}', \mathbf{h}_{\text{lstm}} \right) \\
\mathbf{z}_{\text{encoder}} &= \text{ReTeLU} \left( \mathbf{W}_{\text{fc}} \mathbf{z} + \mathbf{b}_{\text{fc}} \right)
\end{align}
where \( \mathbf{W}_{\text{fc}} \in \mathbb{R}^{(C_{\text{cnn}} + L_{\text{lstm}}) \times D} \) and \( \mathbf{b}_{\text{fc}} \in \mathbb{R}^D \) are the weight matrix and bias for the fully connected layer, respectively.

\subsection{Swarm-Specific Classifier}
The swarm-specific classifier supports the model to the varying number of attack classes for heterogeneous UAV groups. After the shared CNN-LSTM encoder processes the input data, the resulting latent vector \( \mathbf{z}_{\text{encoder}} \in \mathbb{R}^{B \times L_{\text{encoder}}} \) is passed through a multi-layer perceptron (MLP) layer to project the features into a smaller latent space and produce the final class probabilities. The MLP operations are given by:
\begin{align}
\mathbf{y}_{\text{fc1}} &= \mathbf{W}_{\text{fc1}} \mathbf{z}_{\text{encoder}} + \mathbf{b}_{\text{fc1}} \\
\mathbf{y}_{\text{classifier}} &= \text{ReTeLU}(\mathbf{W}_{\text{out}} \mathbf{y}_{\text{fc1}} + \mathbf{b}_{\text{out}})
\end{align}
where \( \mathbf{W}_{\text{fc1}} \), \( \mathbf{b}_{\text{fc1}} \), \( \mathbf{W}_{\text{out}} \), and \( \mathbf{b}_{\text{out}} \) are the weight matrices and biases for the fully connected layers. The softmax function~\cite{wang2018high} ensures that the output is a probability distribution over the attack classes.

\section{Implementation}
In this section, we discuss the four heterogeneous datasets: UAV\_IDS, UKM\_IDS, TLM\_IDS, and the Cyber-Physical dataset, and our unified system model to effectively detect intrusion in UAV swarm networks. To address the challenges of catastrophic forgetting and improve learning efficiency, we incorporate the Elastic Weight Consolidation (EWC) strategy into our federated learning system, transforming it into a federated continuous learning model. This integration allows the system to retain previously learned knowledge while efficiently acquiring new information.

The shared encoder of our proposed encoder-classifier participates directly in federated aggregation, enabling the unified model to detect intrusions across all datasets. Meanwhile, the swarm-specific input layer and local classifiers handle the varying dimensions of each dataset, preserving local knowledge specific to each swarm. This framework ensures fairness without any bias towards specific datasets. It also enables the model to detect intrusions in any new dataset or environment while maintaining the privacy of each swarm's data.  This section provides a detailed description of the datasets, system architecture, federated learning implementation, experimental parameters, and setup.

\subsection{Dataset Descriptions}
Our proposed system model integrates a federated learning framework to combine four heterogeneous datasets: UAV IDS, UKM IDS, TLM IDS, and the Cyber-Physical dataset. The details of these datasets are described in this section and summarized in Table~\ref{tab:dataset_summary}.

\begin{table}[ht]
\centering
\caption{Dataset Summary}
\begin{tabular}{lcccc}
\hline
\textbf{Dataset} & \textbf{Samples} & \textbf{Attributes} & \textbf{Classes} & \textbf{Source} \\ \hline
UAV\_IDS         & 98,736           & 54                  & 2                & UAV Traffic     \\
UKM\_IDS         & 12,887           & 46                  & 9                & Network Logs    \\
TLM\_IDS         & 12,254           & 18                  & 5                & UAV Simulation  \\
Cyber-Physical   & 33,102           & 36                  & 3                & Tello Drone     \\ \hline
\end{tabular}
\label{tab:dataset_summary}
\end{table}

The UAV\_IDS~\cite{UAVIDS2020_Dataset} dataset consists of real-world network traffic data collected from three popular UAV model: DJI Spark, DBPower UDI, and Parrot Bebop. It includes 98,736 unique and up-to-date Wi-Fi traffic log samples, categorized into two classes: Benign and Anomaly. The anomaly class encompasses a broad range of attack vectors. For our experiment, we use the bi-directional flow version of this dataset, which features 54 attributes (+1 label).

The UKM-IDS~\cite{al2021adaptive} dataset collected from real-world network traffic consists of 12,887 samples with 46 features (+2 label), covering nine types of attacks: Normal, TCP flood, Port scanning, ARP poisoning, UDP data flood, Mass HTTP requests, Metasploit exploits, and BeEF HTTP exploits. The dataset's complexity is shown by analyzing its features and classes with rough-set theory and testing it with a dynamic neural network, which demonstrates its higher complexity and relevance compared to other intrusion detection datasets.

The TLM dataset~\cite{yang2023acquisition} was created using a software-in-the-loop simulation setup, where typical UAV failure scenarios are simulated by varying internal physical parameters. It consists of 12,254 samples with 18 attributes (+1 label) and five class labels: Benign, RC failure, GPS failure, ACC failure, and Engine failure. A quadrotor UAV is used to simulate common UAV anomalies, such as engine failures, accelerometer malfunction, and remote control issues for data collection.

The Cyber-Physical Dataset~\cite{puccetti2024rospace} integrates both cyber and physical attributes to create an intrusion detection dataset, featuring five different attack classes. The data was collected using the DJI Tello EDU drone. Due to differing data dimensions across the attack classes, we use a minimized version of the dataset, which contains 33,102 samples, 36 features, and one label with three classes: Benign, Replay, and DoS.

\subsection{System Architecture}
\begin{figure}[ht]
    \centering
    \includegraphics[width=1\linewidth]{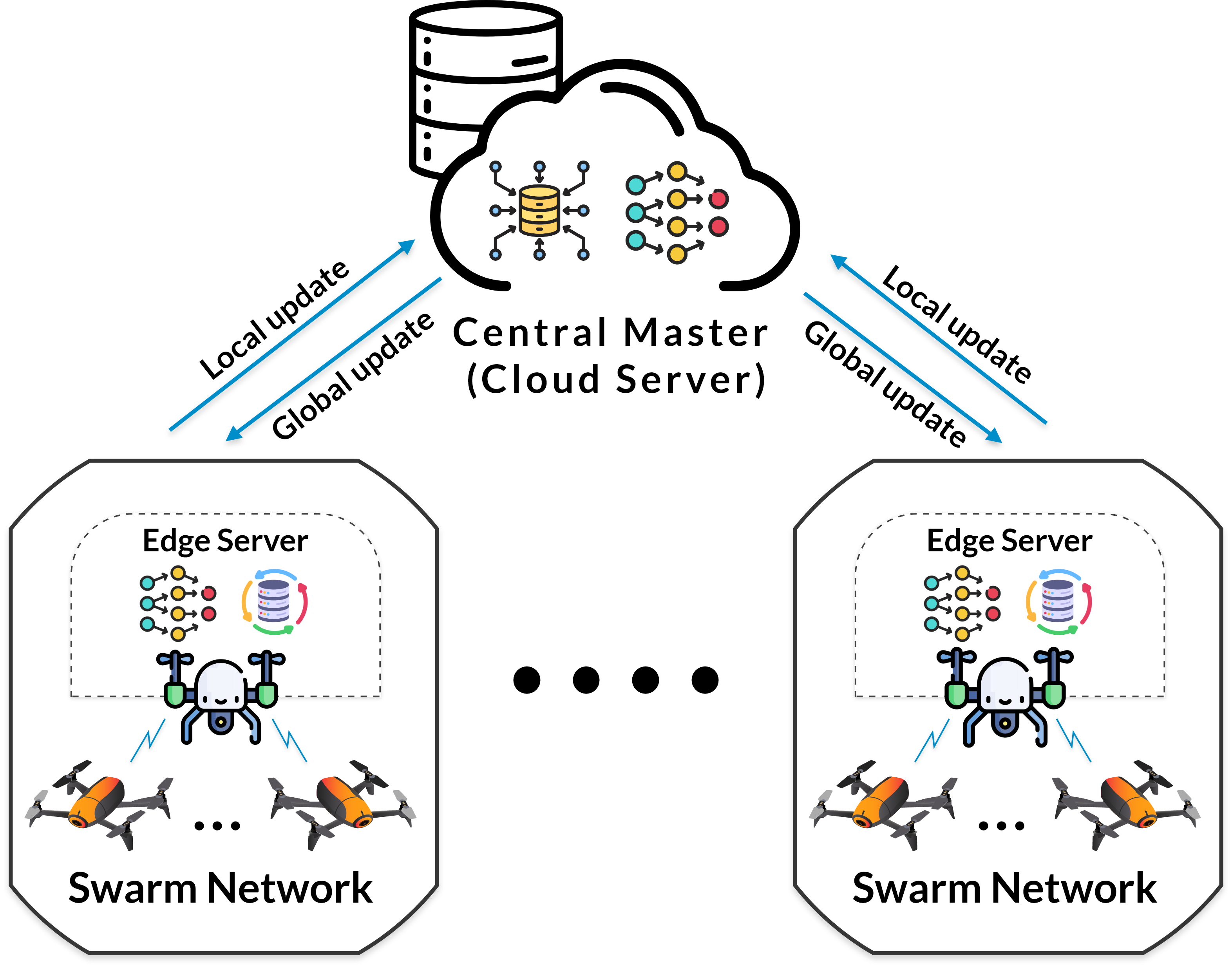}
    \caption{Federated learning based system architecture}
\label{fig:system_architecture}.
\end{figure}

Our proposed system architecture integrates four separate UAV swarm networks, each corresponding to one of our heterogeneous datasets, to perform decentralized intrusion detection using Federated Learning. Each swarm consists of several leaf UAV nodes and an edge server, with the edge server acting as a local master. The edge server maintains a copy of the Encoder-Classifier model, aggregates data from the UAV nodes, and performs local anomaly detection. A central cloud server coordinates the aggregation of model updates from all edge servers to maintain a global model. While all servers share the same Encoder-Classifier framework, the shared encoder weights are synchronized across the system, with separate input layers and classifiers specific to each swarm. The proposed system architecture is depicted in Fig.~\ref{fig:system_architecture}.

To train the global model without sharing raw data, we adopt the Federated Averaging (FedAvg) algorithm \cite{ rahman2025fair}. In each round, a subset of clients (UAV swarms) trains their local models, and the local model updates are averaged to update the global model. The client-specific update \( w_i^{(r)} \) is computed as:
\begin{align}
w_i^{(r)} = w_i^{(r-1)} - \eta \nabla L(w_i^{(r-1)}, D_i)
\end{align}

where \( L(w_i, D_i) \) is the local loss and \( \eta \) is the learning rate. The global model is updated by averaging the local updates:
\begin{align}
w^{(r)} = \frac{1}{|S_r|} \sum_{i \in S_r} w_i^{(r)}
\end{align}
where \( S_r \) is the set of selected clients in round \( r \).

Elastic Weight Consolidation (EWC) \cite{aslam2025cel} is used to mitigate catastrophic forgetting when training across multiple heterogeneous tasks, as well as obtain the continuous learning capability \cite{maschler2020continual}. The EWC penalty term is added to the local loss:
\begin{align}
\mathcal{L}(w) = \mathcal{L}_{\text{local}}(w, D_i) + \mathcal{L}_{\text{EWC}}(w)
\end{align}
where \( \mathcal{L}_{\text{EWC}}(w) = \sum_{i} \lambda_i (w_i - \hat{w}_i)^2 \) and \( \hat{w}_i \) represents the optimal parameter from the previous task. The step-by-step process of the FedAvg loop with EWC integration is given in Algorithm~\ref{alg:fedavg}.

\begin{algorithm}[ht]
\caption{Federated Learning for Intrusion Detection}
\label{alg:fedavg}
\begin{small}
\begin{algorithmic}[1]
\STATE \textbf{Input:} \( \theta_{\text{enc}}, \theta_{\text{clf}}, D_i, \lambda_i \)
\STATE \textbf{Output:} Global model parameters \( \theta_{\text{global}} \)
\STATE Initialize global parameters \( \theta_{\text{global}}^{(0)} \)
\FOR{each round \( r = 1, \dots, 50 \)}
    \STATE Select a subset \( S_r \subset K \) of clients
    \STATE Send \( \theta_{\text{global}}^{(r-1)} \) to clients in \( S_r \)
    \FOR{each client \( k \in S_r \)}
        \STATE Perform local training and compute update \( w_k^{(r)} \) using:
        \[
        w_k^{(r)} = w_k^{(r-1)} - \eta \nabla L(w_k^{(r-1)}, D_k) + \lambda \mathcal{L}_{\text{EWC}}(w_k^{(r-1)}, \theta_k)
        \]
    \ENDFOR
    \STATE Aggregate updates:
    \[
    \theta_{\text{global}}^{(r)} = \frac{1}{|S_r|} \sum_{k \in S_r} w_k^{(r)}
    \]
\ENDFOR
\end{algorithmic}
\end{small}
\end{algorithm}


The cloud server aggregates model updates from edge servers after each round. The global model is then distributed to the edge servers for real-time anomaly detection, ensuring privacy-preserving training and continuous improvement through federated updates and EWC regularization.

\subsection{Experimental Parameters}
In this study, we use the Flower framework\footnote{https://flower.ai/} for federated learning simulations and PyTorch\footnote{https://pytorch.org/} for deep learning model development. The experiments were conducted on a Windows 11 PC with an Intel Core i7-7500U 2.7GHz Processor, an NVIDIA GeForce 940MX 4GB DDR3 GPU, and 8GB DDR4 RAM. The dataset was split into a training set (80\%) and a test set (20\%). The configurations and characteristics of the test environment are summarized in Tables \ref{tab:model_summary}, \ref{tab:training_parameters}, and \ref{tab:client_config}.

\begin{table}[ht]
\centering
\caption{Encoder-Classifier Model Summary}
\begin{tabular}{lll}
\hline
\textbf{Layer Name} & \textbf{Input Dimension} & \textbf{Output Dimension} \\ \hline
Input Layer          & \( \text{\textit{Dynamic}} \)       & \( 128 \)  \\ 
CNN Layer            & \( 128 \times 1 \)            & \( 16 \times 1 \) \\ 
BatchNorm1d          & \( 16 \times 1 \)             & \( 16 \times 1 \)  \\ 
MaxPool1d            & \( 16 \times 1 \)             & \( 16 \times 1 \)  \\
LSTM Layer           & \( 16 \times 1 \)             & \( 128 \times 1 \) \\ 
Dense      & \( 144 \)                     & \( 64 \) \\
Classifier           & \( 64 \)                      & \(\text{\textit{Dynamic}} \)  \\ \hline
\end{tabular}
\label{tab:model_summary}
\end{table}

\begin{table}[ht]
\centering
\caption{Model Training Parameters}
\begin{tabular}{p{5.5cm}c}
\hline
\textbf{Parameter} & \textbf{Value} \\ \hline
Batch Size         & 32 \\
Epochs             & 1   \\
Rounds             & 50 \\
Learning Rate      & 0.001 \\
Lambda EWC (Regularization) & 0.4    \\ \hline
\end{tabular}
\label{tab:training_parameters}
\end{table}

\begin{table}[ht]
\centering
\caption{FL Client Configuration Parameters}
\begin{tabular}{p{5.5cm}c}
\hline
\textbf{Parameter} & \textbf{Value} \\ \hline
Num\_UAV\_Client & 4 \\
Min\_fit\_UAV\_clients & 4 \\
Min\_evaluate\_UAV\_clients & 4 \\
Min\_available\_UAV\_clients & 4 \\ \hline
\end{tabular}
\label{tab:client_config}
\end{table}

We employ the Cross-Entropy Loss function \cite{mao2023cross} to optimize the model, as it effectively measures how far the predicted probability distribution \( p \) deviates from the actual label distribution \( y \). The loss is computed as:

\begin{align}
    \mathcal{L}_{\text{CE}} = - \sum_{c=1}^{C} y_c \log(p_c)
\end{align}

where \( C \) represents the number of classes, \( y_c \) is the true label, and \( p_c \) is the predicted probability for class \( c \). This loss function is effective in classification tasks, driving the model towards accurate class predictions.

Given the federated learning setup, which involves multi-party collaboration, distributed computation, and data privacy, the model can efficiently train on heterogeneous UAV swarm data while safeguarding privacy, even with limited computational resources.

\section{Results and Analysis}
In this section, we conduct a comprehensive evaluation of the proposed system in terms of classification accuracy, computational efficiency, and communication overhead.

\subsection{Classification Performance}
\begin{figure*}[h!]
    \centering
    
    \begin{subfigure}[b]{0.4\linewidth}
        \centering
        \includegraphics[width=\textwidth]{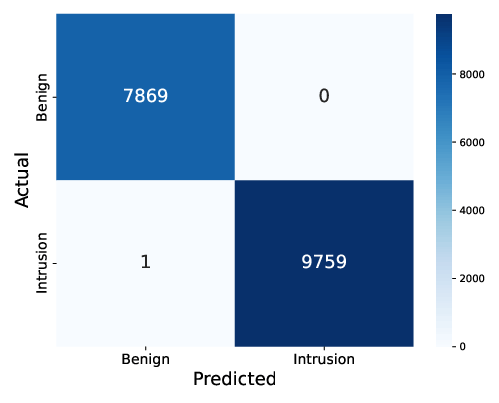}
        \caption{UAV\_IDS}
        \label{fig:UAV_IDS}
    \end{subfigure}
    \begin{subfigure}[b]{0.4\linewidth}
        \centering
        \includegraphics[width=\textwidth]{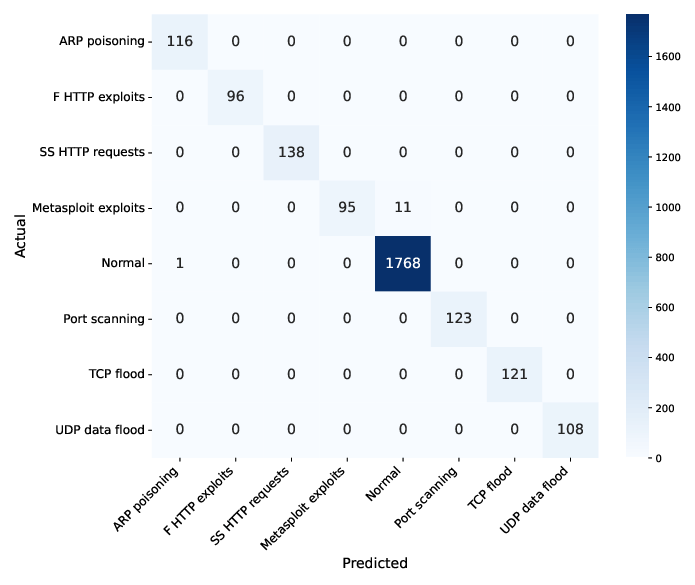}
        \caption{UKM\_IDS}
        \label{fig:UKM_IDS}
    \end{subfigure}
    
    \vskip\baselineskip
    
    \begin{subfigure}[b]{0.4\linewidth}
        \centering
        \includegraphics[width=\textwidth]{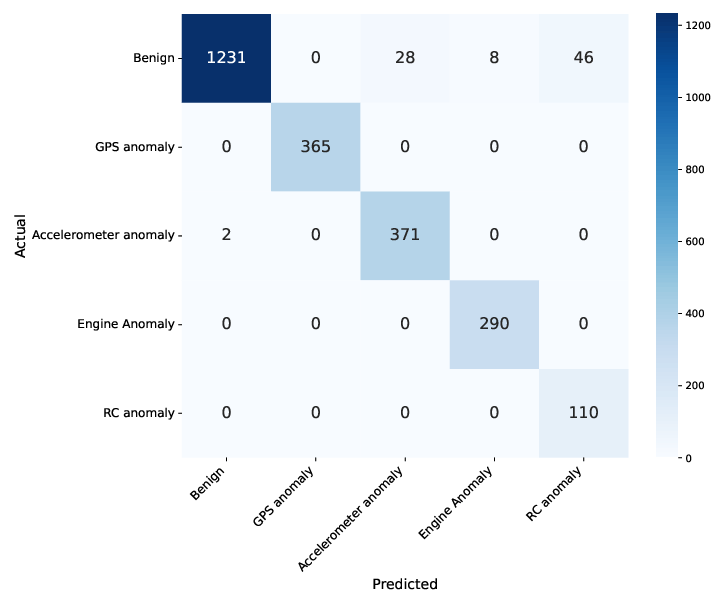}
        \caption{TLM\_IDS}
        \label{fig:TLM_IDS}
    \end{subfigure}
    \begin{subfigure}[b]{0.4\linewidth}
        \centering
        \includegraphics[width=\textwidth]{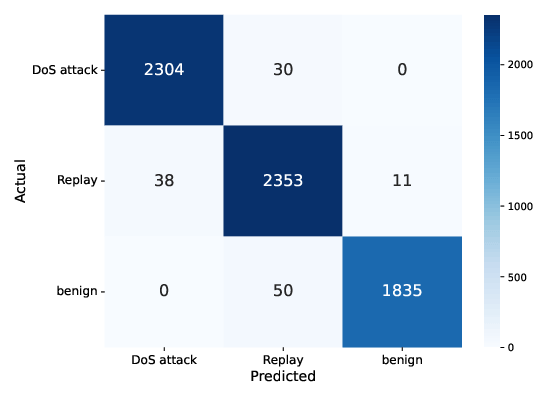}
        \caption{Cyber\_Physical}
        \label{fig:Cyber_Physical}
    \end{subfigure}
    
    \caption{Confusion Matrix of Four Heterogeneous Swarm}
    \label{fig:confusion_matrices}
\end{figure*}

To evaluate the effectiveness of our proposed system, we utilize a multiclass confusion matrix analysis on the testing datasets. The confusion matrix reveals the distribution and recognition patterns relative to the true labels, providing insights into the model's performance. It is important to note that this evaluation is conducted after post-federated learning training, during which identical models (with shared weights) are tested on four distinct swarms corresponding to our four heterogeneous datasets. The evaluation results, derived from the confusion matrices, are shown in Fig.~\ref{fig:confusion_matrices}.

Fig.~\ref{fig:confusion_matrices} shows the multiclass confusion matrices for the four swarm datasets. In all cases, most predictions lie along the diagonal, indicating that the model accurately classifies both benign and attack samples. The UAV\_IDS dataset shows near-perfect performance with only one error. UKM\_IDS also performs strongly, with very few misclassifications across multiple attack types. For TLM\_IDS, the model correctly identifies most classes, though there are a few confusions between similar anomaly types. On the Cyber\_Physical dataset, the model shows high accuracy in detecting DoS and replay attacks, with only minor errors in classifying benign samples. Overall, the model demonstrates reliable and consistent performance across all datasets.

Moreover, additional performance evaluation measures, including the Detection Accuracy, Recall, F1 Score, Detection Precision, Detection Error Rate~\cite{hadi2024real} are calculated using the confusion matrix parameters, which are presented in Table~\ref{tab:evaluation_scores}.

\begin{table}[ht]
\centering
\caption{Evaluation Scores of Four Heterogeneous Swarm}
\begin{tabular}{p{1.4cm}cccc|l}
\hline
\textbf{Dataset} & \textbf{Accuracy} & \textbf{Recall} & \textbf{F1 Score} & \textbf{Precision} & \textbf{Error} \\
\hline
UAV\_IDS         & 99.99\% & 99.99\%  & 99.99\% & 99.99\% & 0.01\% \\
UKM\_IDS         & 99.46\% & 99.61\%  & 99.03\% & 99.62\% & 0.54\% \\
TLM\_IDS         & 96.85\% & 98.64\%  & 94.83\% & 92.13\% & 3.43\% \\
CyberPhysical  & 98.05\% & 98.00\%  & 98.08\% & 98.16\% & 1.95\% \\
\hline
\end{tabular}
\label{tab:evaluation_scores}
\end{table}

\begin{figure}[h!]
    \centering
    \includegraphics[width=1\linewidth]{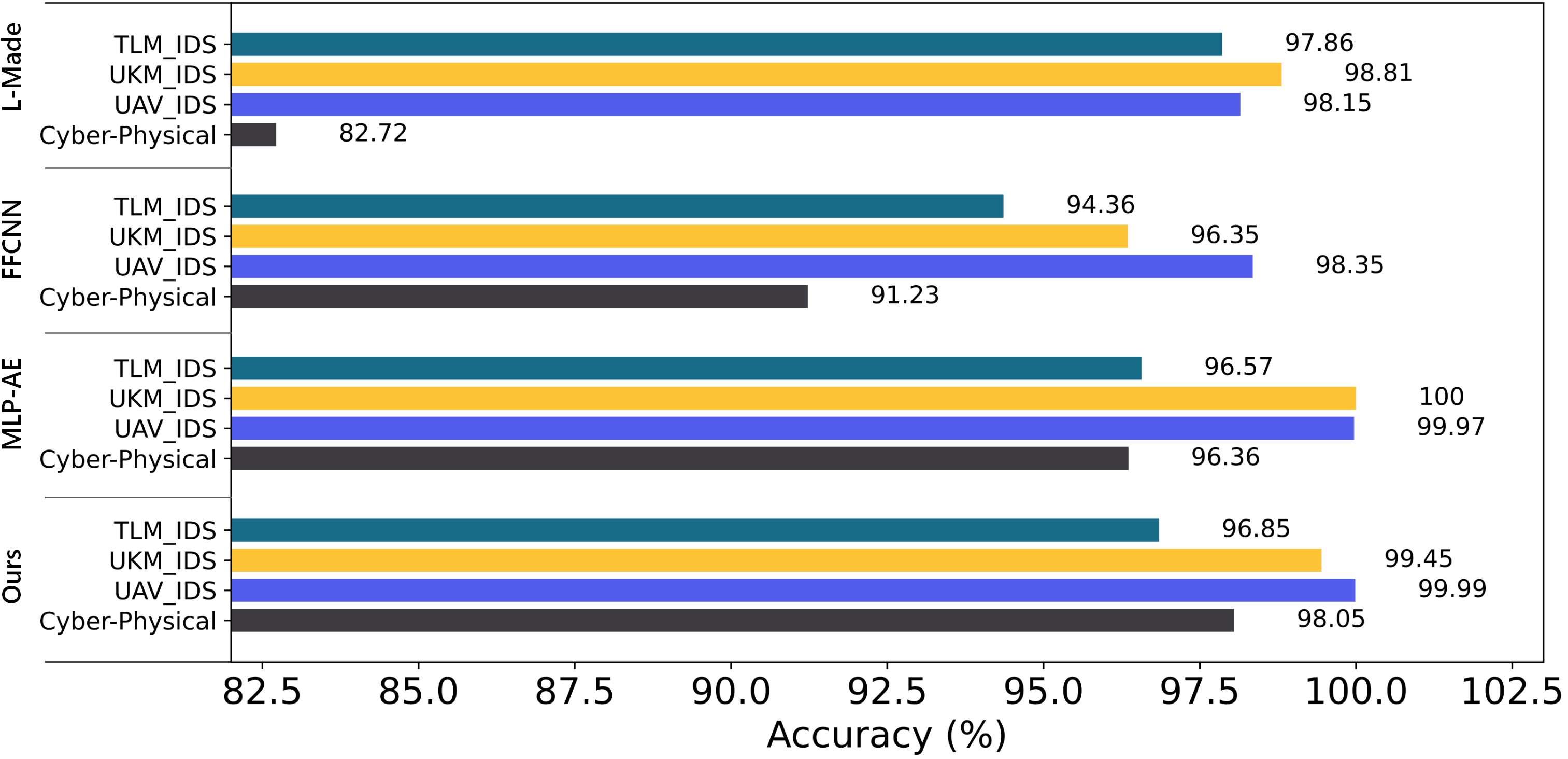}
    \caption{Accuracy Comparison across different dataset}
\label{fig:comparison_bar}
\end{figure}

To demonstrate the improvements, we conducted experiments using several existing benchmarks, including the Multi-modal Denoising Auto-encoder (L-MADE)~\cite{lu2024swarm}, Feed Forward CNN (FFCNN)~\cite{hadi2024real}, MLP Auto-encoder (MLP-AE)~\cite{yashwanth2024network}, and our CNN-LSTM-based Encoder-Classifier model. The results presented in Table~\ref{tab:comparison} and Fig.~\ref{fig:comparison_bar} highlight the performance differences across these models. 

\begin{table}[h!]
\centering
\caption{Performance Comparison Across Various Models}
\begin{tabular}{p{1.3cm}lcccc}
\hline
\multicolumn{1}{l}{\textbf{Datasets}} & \textbf{Model} & \textbf{Accuracy} & \textbf{Recall} & \textbf{F1} & \textbf{Precision} \\ 
\hline
\multirow{4}{*}{UAV\_IDS} & L-MADE   & 98.15\% & 97.84\% & 98.10\% & 97.84\%  \\ 
                          & FFCNN    & 98.35\% & 99.87\% & 99.00\% & 99.64\%  \\ 
                          & MLP-AE   & 99.97\% & 99.97\% & 99.99\% & 99.97\%  \\ 
                          & \textbf{Ours}     & \textbf{99.99\%} & \textbf{99.99\%} & \textbf{99.99\%} & \textbf{99.99\%}  \\ 
                          \hline
\multirow{4}{*}{UKM\_IDS} & L-MADE   & 98.81\% & 98.67\% & 99.23\% & 98.76\%  \\ 
                          & FFCNN    & 96.35\% & 98.88\% & 99.00\% & 99.35\%  \\ 
                          & MLP-AE   & 100\% & 100\% & 100\% & 100\%  \\ 
                          & \textbf{Ours}     & \textbf{99.46\%} & \textbf{99.61\%} & \textbf{99.03\%} & \textbf{98.62\%}  \\
                          \hline
\multirow{4}{*}{TLM\_IDS} & L-MADE   & 97.86\% & 97.60\% & 98.86\% & 98.52\%  \\ 
                          & FFCNN    & 94.36\% & 79.70\% & 84.00\% & 91.40\%  \\ 
                          & MLP-AE   & 96.57\% & 96.93\% & 95.12\% & 90.94\%  \\ 
                          & \textbf{Ours}     & \textbf{96.85\%} & \textbf{98.64\%} & \textbf{94.83\%} & \textbf{92.13\%}  \\
                          \hline
\multirow{4}{*}{CyberPhysical} & L-MADE & 82.72\% & 83.80\% & 82.72\% & 83.06\%  \\ 
                               & FFCNN  & 91.83\% & 88.56\% & 92.67\% & 94.94\%  \\ 
                               & MLP-AE & 96.36\% & 96.36\% & 96.49\% & 96.80\%  \\ 
                          & \textbf{Ours}     & \textbf{98.05\%} & \textbf{98.00\%} & \textbf{98.08} & \textbf{98.16\%}  \\
\hline
\end{tabular}
\label{tab:comparison}
\end{table}

Table~\ref{tab:comparison} shows that our approach consistently performs at or near the best across all datasets. It clearly outperforms the baselines on UAV-IDS and CyberPhysical, and performs competitively on UKM-IDS and TLM-IDS. Most notably, our model stands out for its stable and strong performance across all datasets, highlighting its robustness and efficiency.

\subsection{Computational and Communication Performance}
In addition to achieving high detection accuracy, our proposed system is optimized to be lightweight and suitable for deployment in real UAV swarms. This section evaluates the system’s training cost, inference delay, and communication overhead throughout the federated learning process.

\begin{figure}[ht]
    \centering
    \begin{subfigure}[b]{1\linewidth}
        \centering
        \includegraphics[width=\linewidth]{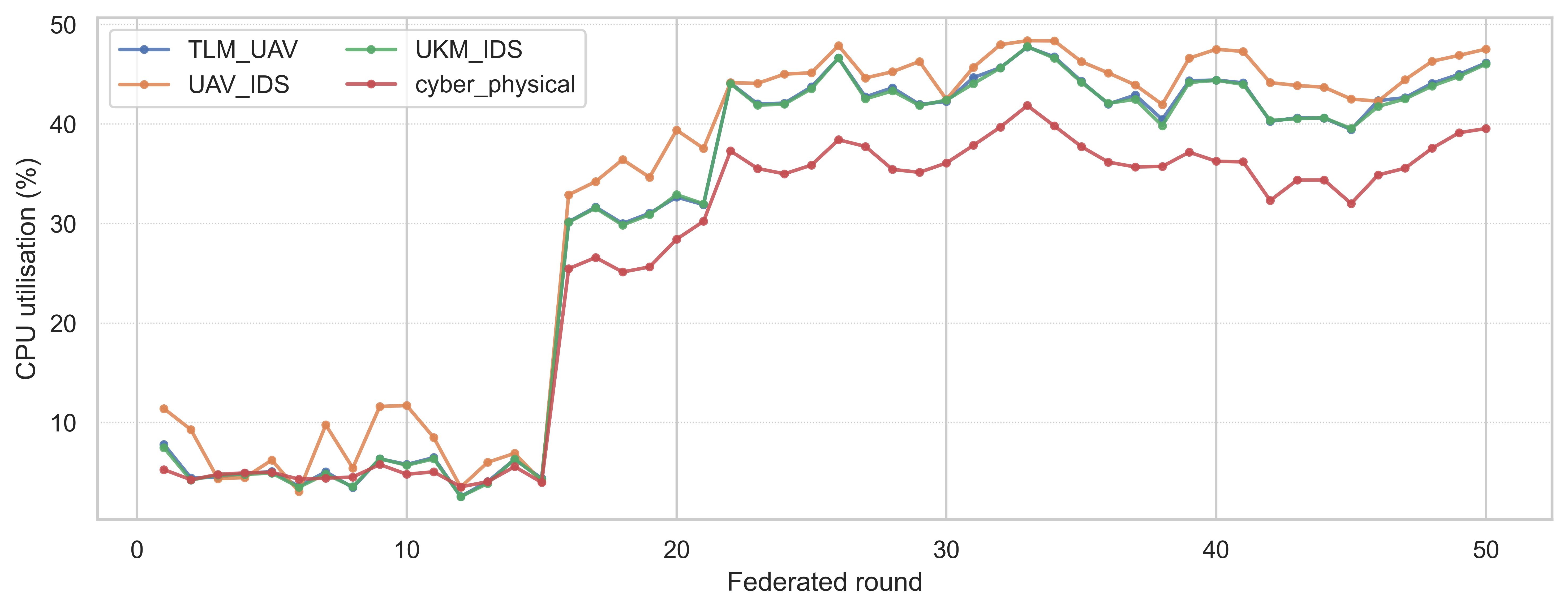}
        \caption{CPU utilization}
        \label{fig:cpu_util_line}
    \end{subfigure}
    \vskip\baselineskip
    \begin{subfigure}[b]{1\linewidth}
        \centering
        \includegraphics[width=\linewidth]{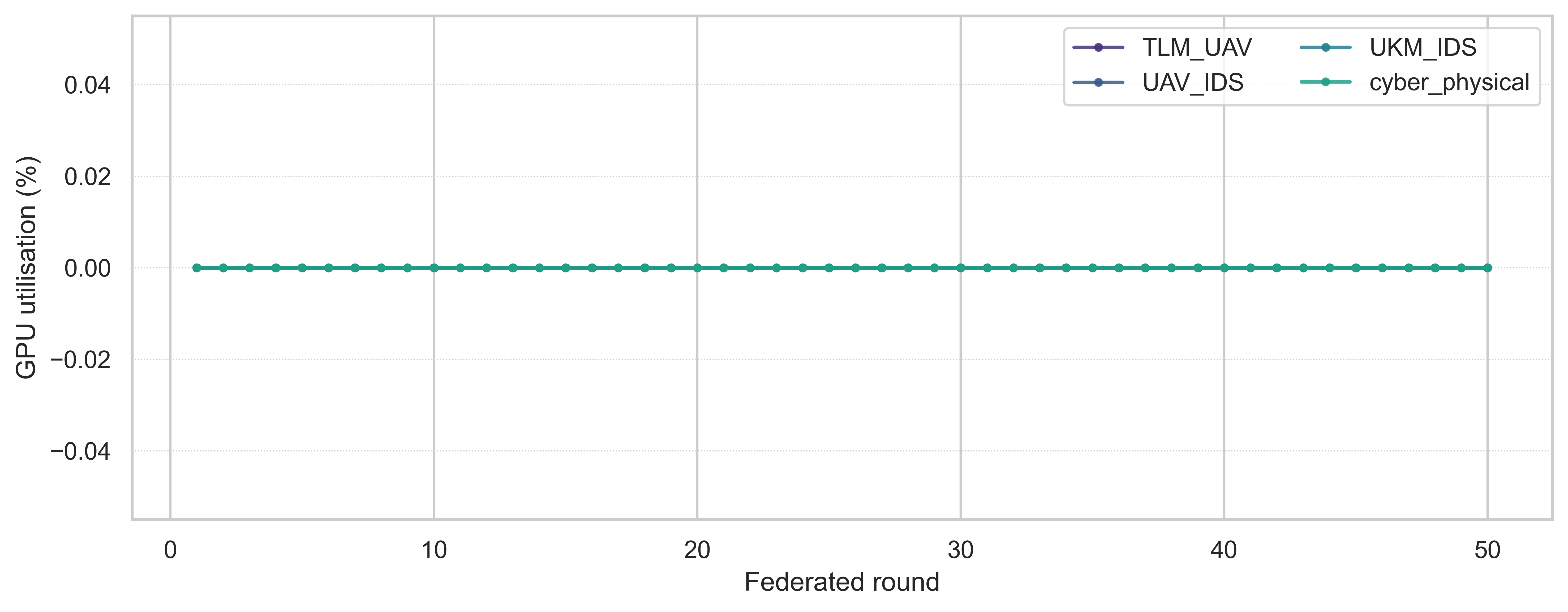}
        \caption{GPU utilization}
        \label{fig:gpu_util_line}
    \end{subfigure}
    \vskip\baselineskip
    \begin{subfigure}[b]{1\linewidth}
        \centering
        \includegraphics[width=\linewidth]{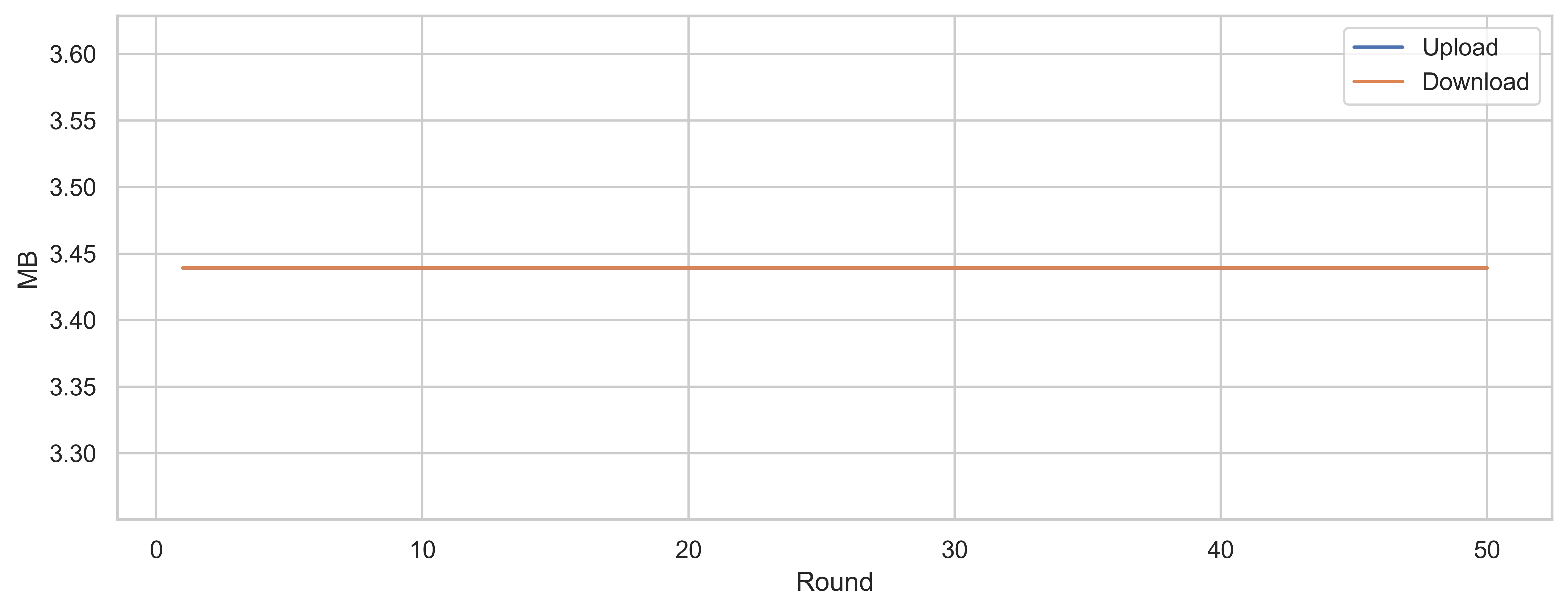}
        \caption{Bandwidth per round}
        \label{fig:bw_round}
    \end{subfigure}
    \caption{System resource utilization during 50 federated rounds.}
    \label{fig:comp_system}
\end{figure}

Fig.~\ref{fig:comp_system} shows three key system-level metrics: CPU usage, GPU usage, and communication bandwidth per round. From Fig.~\ref{fig:cpu_util_line}, we observe that CPU usage gradually rises during training and stabilizes around 40–50\% for most clients. This confirms that the model trains well even on basic UAV-grade processors. Fig.~\ref{fig:gpu_util_line} confirms that GPU utilization is almost zero, which means our system does not rely on GPUs and can easily run on edge devices without accelerators. In Fig.~\ref{fig:bw_round}, we see that both upload and download sizes remain fixed at around 3.45 MB, showing low communication cost and consistency in each round.

\begin{figure}[ht]
    \centering
    \includegraphics[width=1\linewidth]{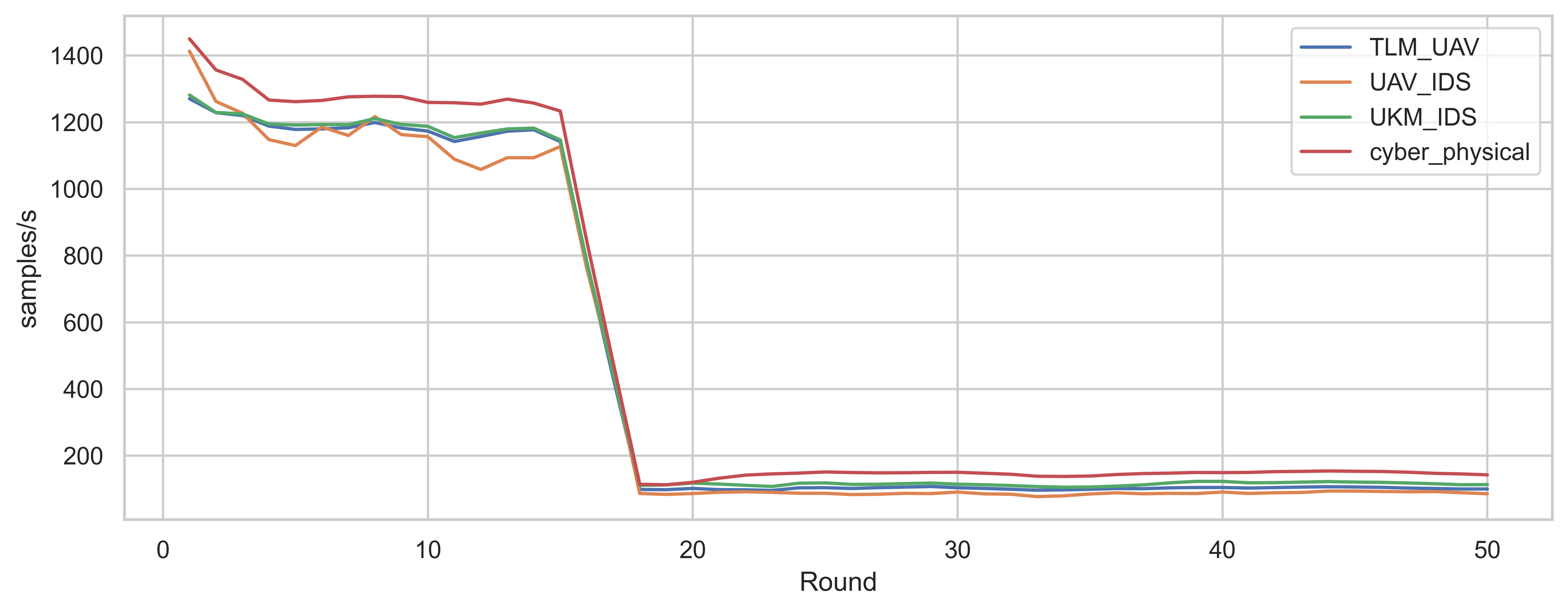}
    \caption{Client-side throughput across federated training rounds.}
    \label{fig:throughput_line}
\end{figure}

To understand how fast each client can process data during training, we measure the number of samples trained per second in every round. As shown in Fig.~\ref{fig:throughput_line}, clients start with high throughput (1200–1400 samples/s) and later stabilize around 100–150 samples/s. This steady pattern confirms that training remains smooth and manageable on all clients, even when hardware capabilities differ.

\begin{figure}[ht]
    \centering
    \includegraphics[width=1\linewidth]{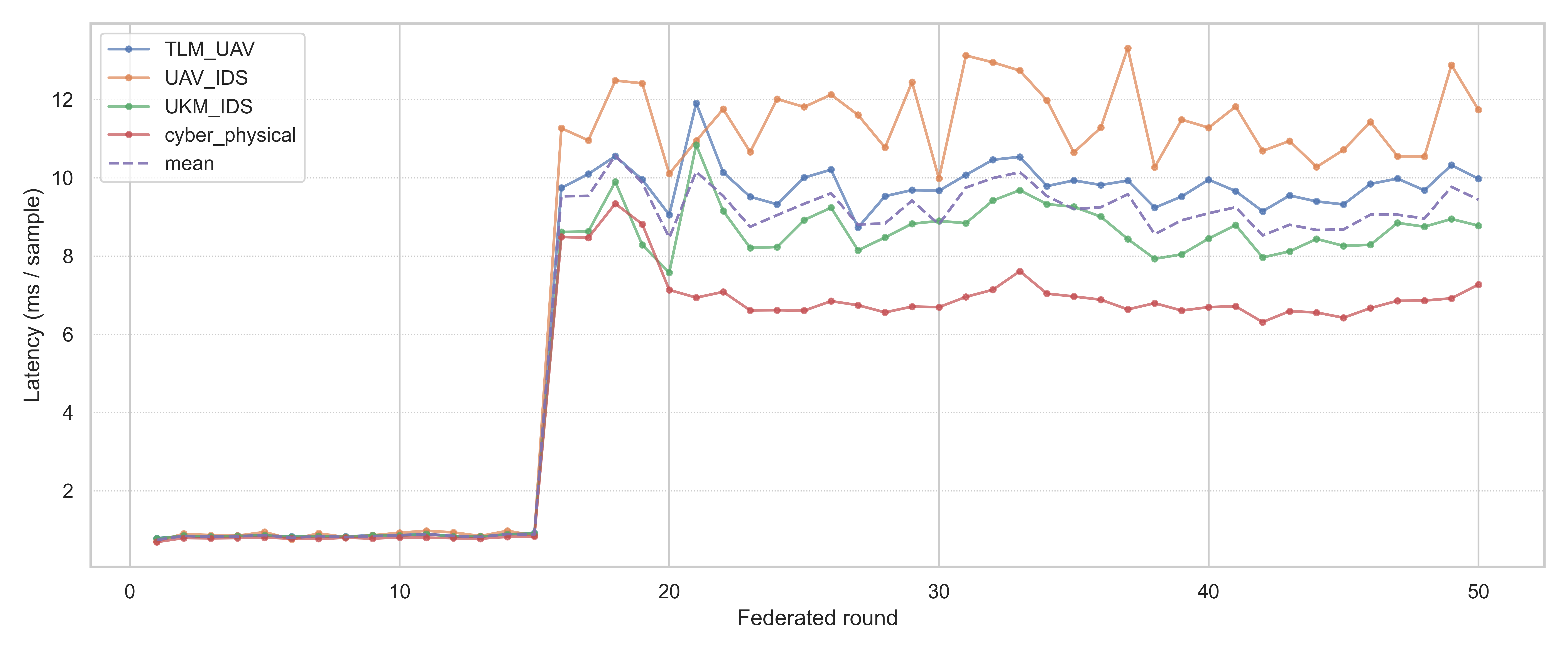}
    \caption{Per-sample inference latency (lower is better).}
    \label{fig:latency_line}
\end{figure}

It is crucial to detect intrusion in real-time or near-real-time to ensure UAV safety in a dynamic environment. Fig.~\ref{fig:latency_line} shows the average inference latency per sample for each client. The latency stays well below 15 milliseconds per sample, with most clients operating between 8–12 ms. This observation states that our model responds quickly and is suitable for real-world deployment.

\begin{figure}[ht]
    \centering
    \includegraphics[width=1\linewidth]{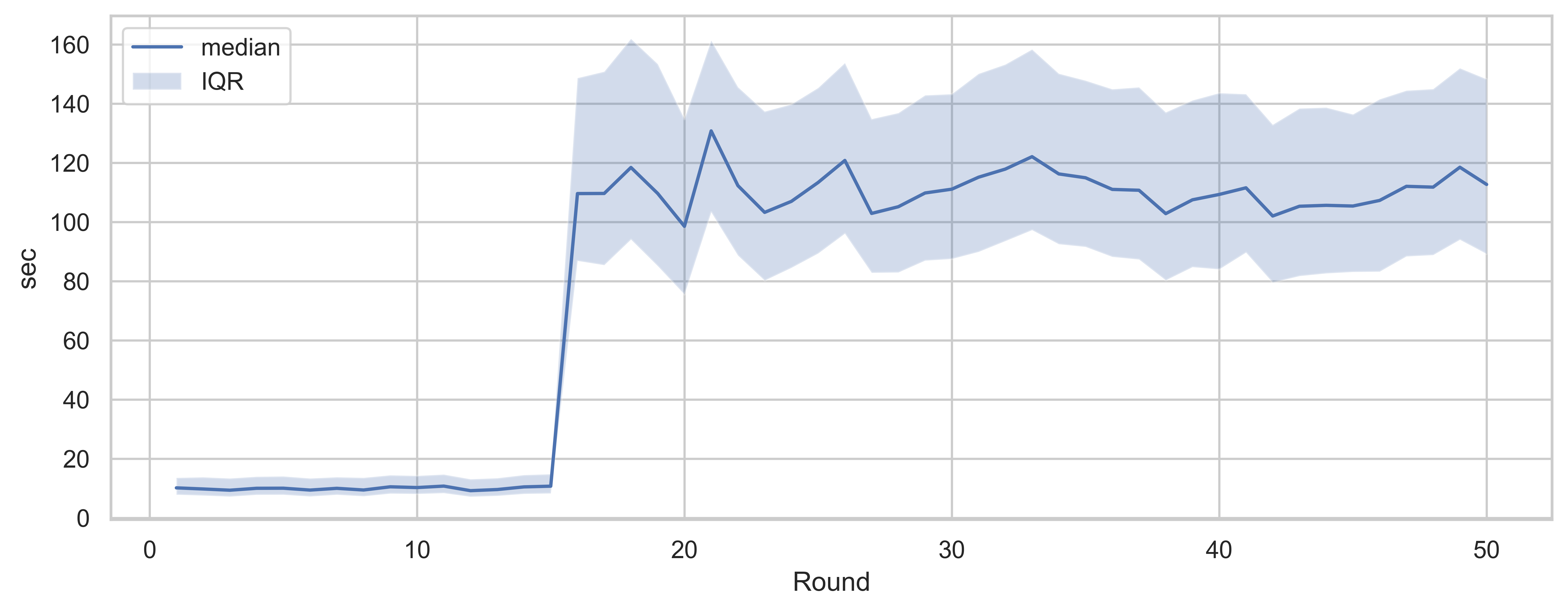}
    \caption{Client training time per round (median and IQR).}
    \label{fig:fit_time_line}
\end{figure}

Fig.~\ref{fig:fit_time_line} tracks the total time each client takes to complete training in each round. The median fit-time remains under 150 seconds throughout the 50 rounds. This trend is consistent and predictable, which helps when coordinating training schedules in real UAV fleets.

\begin{table}[ht]
\centering
\caption{System Profiling Summary}
\begin{tabular}{lcc>{\centering\arraybackslash}p{2cm}}
\hline
\textbf{Metric} & \textbf{CPU} & \textbf{Quant. CPU} & \textbf{GPU} \\
\hline
Latency (ms/sample)     & 2.12  & 2.77  & 3.49 \\
CPU Memory (MB)         & 831.04 & 833.71 & 789.60 \\
GPU Memory (MB)         & 8.52  & 8.52  & 9.47 \\
Power Usage (Watt)      & --    & --    & 9.78 \\
\hline
\\
\hline
\multicolumn{3}{l}{Model FLOPs}          & 236,736 \\
\multicolumn{3}{l}{Trainable Params}     & 231,541 \\
\multicolumn{3}{l}{Model Size (bytes)}   & 955,959 \\
\hline
\end{tabular}
\label{tab:sys_profile}
\end{table}

To give a more detailed picture, we also collect static system profiling information during a test inference run. Table~\ref{tab:sys_profile} summarizes key statistics such as model size, Floating-point operations per second (FLOPs), and resource consumption on CPU and GPU.

\begin{figure}[h!]
    \centering
    \begin{subfigure}[b]{0.48\linewidth}
        \centering
        \includegraphics[width=\textwidth]{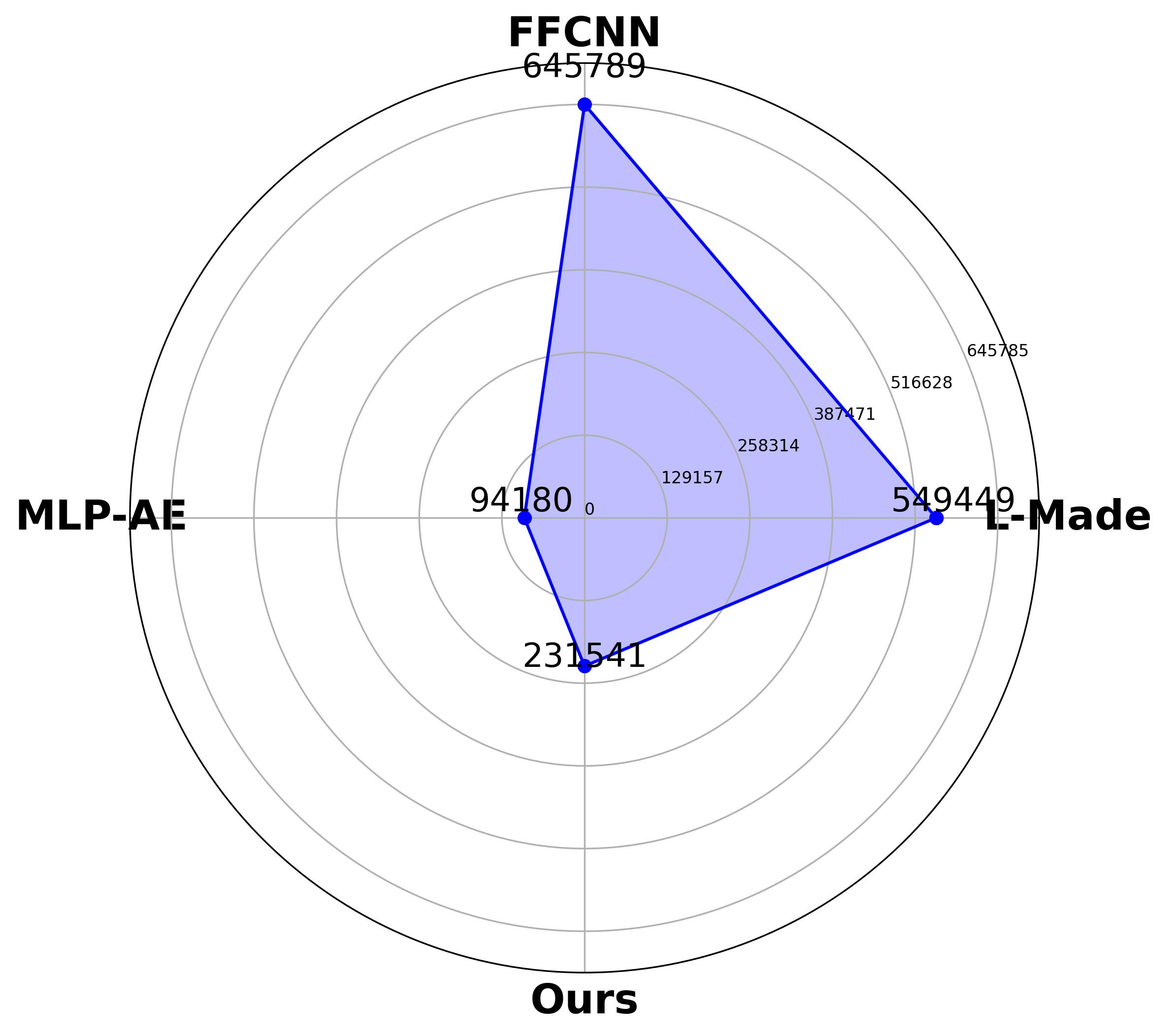}
        \caption{Trainable parameter count}
        \label{fig:model_params_radar}
    \end{subfigure}
    \hfill
    \begin{subfigure}[b]{0.48\linewidth}
        \centering
        \includegraphics[width=\textwidth]{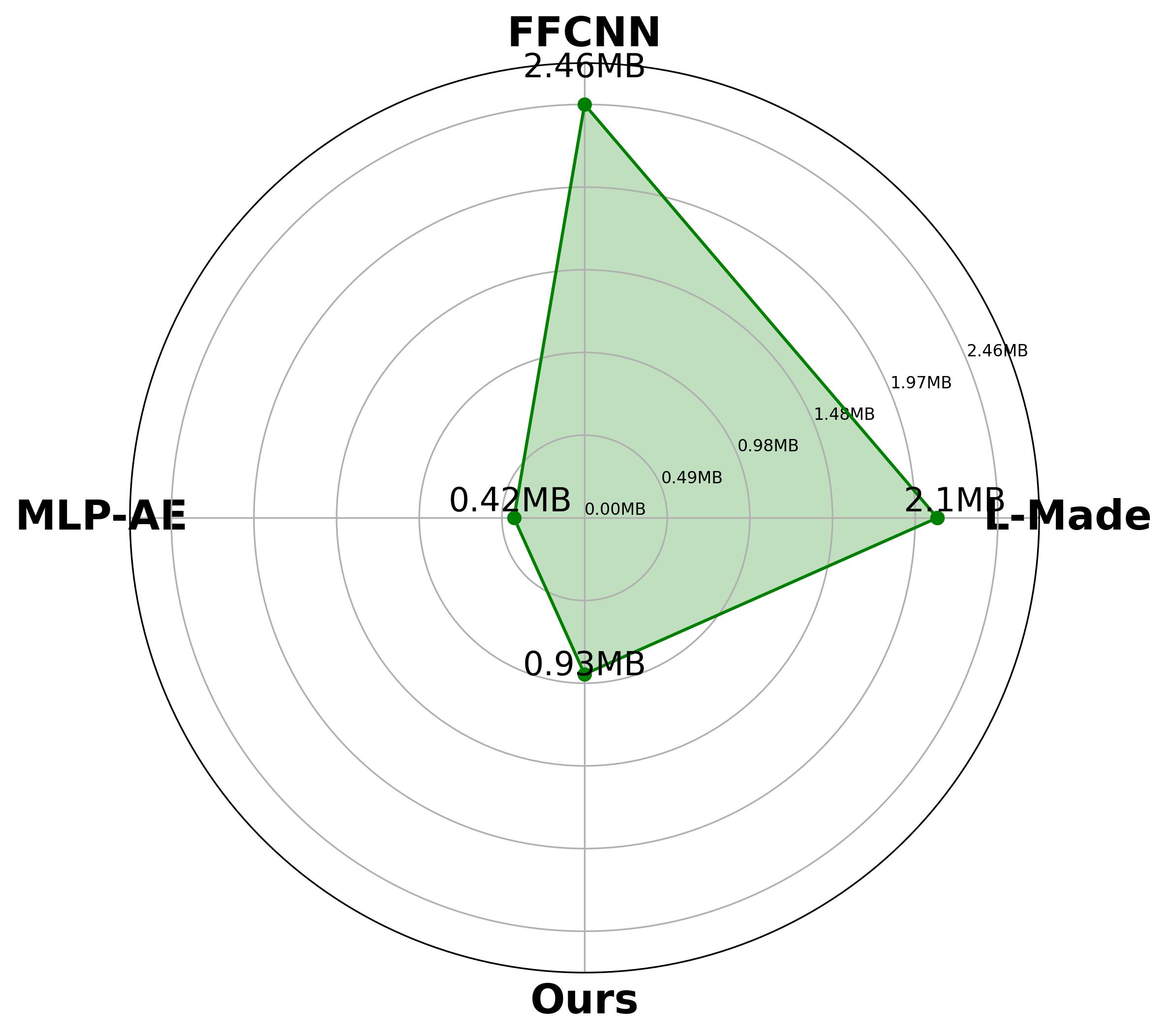}
        \caption{Model size in MB}
        \label{fig:model_size_radar}
    \end{subfigure}
    \caption{Model weight comparison across different architectures.}
    \label{fig:model_weight_comparison}
\end{figure}

Finally, Fig.~\ref{fig:model_weight_comparison} compares our model’s weight and parameter footprint with existing works. The radar plots confirm that our method achieves excellent accuracy while staying compact in size, making it ideal for memory-constrained UAV boards.

Our proposed system model requires low CPU and GPU power, uses minimal bandwidth, and finishes training quickly. These evaluation metrics demonstrate the effectiveness and robustness of the proposed IDS system for real-world deployment.



\section{Conclusion}
In the emerging field of intrusion detection within UAV swarm networks, many existing solutions excel in controlled lab environments but struggle to perform effectively in real-world deployments. These challenges often arise from relying on biases in specific datasets or configurations with limited intrusion patterns, centralized structure, and high computational costs. While some solutions address specific issues such as privacy preservation, noise reduction, and false alarm minimization, they lack a comprehensive and unified framework to tackle the existing research gaps. In contrast, we introduce a unified, decentralized, and lightweight system framework to handle heterogeneous data across UAV swarms effectively. The proposed system model ensures unbiased detection, preserves privacy, reduces false alarms due to noise, supports continuous learning without forgetting previous knowledge, and enhances the detection accuracy of new attack patterns. The performance evaluation of our proposed system model demonstrates robust and competitive performance compared to leading benchmarks. By integrating federated learning with lightweight design, our system not only addresses current challenges but also paves the way for Advanced General Intelligence (AGI) in intrusion detection.

\bibliographystyle{IEEEtran}
\bibliography{biblography}

\end{document}